\input phyzzx
\sequentialequations
\overfullrule=0pt
\tolerance=5000
\nopubblock
\twelvepoint
 
\line{\hfill }
\line{\hfill PUPT/1614, IASSNS 96/35}
\line{\hfill hep-th/9604134 }
\line{\hfill April 1996}

\titlepage
\title{Classical Hair in String Theory I: General Formulation}

\author{Finn Larsen\foot{Research supported in part by Danish National
Science Foundation. larsen@puhep1.princeton.edu}}
\centerline{{\it Department of Physics }}
\centerline{{\it Joseph Henry Laboratories }}
\centerline{{\it Princeton University }}
\centerline{{\it Princeton, N.J. 08544 }}
\vskip .2cm
\author{Frank Wilczek\foot{Research supported in part by DOE grant
DE-FG02-90ER40542.~~~wilczek@sns.ias.edu}}
\vskip.2cm
\centerline{{\it School of Natural Sciences}}

\centerline{{\it Institute for Advanced Study}}
\centerline{{\it Olden Lane}}
\centerline{{\it Princeton, N.J. 08540}}
 
\endpage
 
\abstract{We discuss why classical hair is desirable for the
description of black holes, and show that it arises generically in
a wide class of field theories involving extra dimensions.  We develop
the canonical formalism for theories with the matter content that
arises in string theory.  General covariance and 
duality are used to determine the form of
surface terms. We derive an effective theory (reduced Hamiltonian) 
for the hair in terms of horizon variables.
Solution of the constraints expresses these variables in terms of hair
accessible to an observer at infinity.  We
exhibit some general properties of the resulting theory, including a 
formal identification of the temperature and entropy.  The 
Cveti\v{c}-Youm dyon is described in some detail, as an important example.}
 
\endpage

\REF\wald{R. Wald, {\it General Relativity} University of Chicago press,
Chicago (1984).}
\REF\nohair{R. H. Price, Phys. Rev. {\bf D5} (1972) 2419, 2439 .}
\REF\hairrefs{B. A. Campbell, N. Kaloper, and K. A. Olive,
Phys. Lett. {\bf B263} (1991),
J. Ellis, N. Mavromatos, and D. V. Nanopoulos, 
Phys. Lett. {\bf B294} 229 (1992).}
\REF\qhair{S. Coleman, J. Preskill, and F. Wilczek,
Nucl. Phys. {\bf B378} 175 (1992). }
\REF\sen{A. Sen. Mod. Phys. Lett. {\bf A10} 2081 (1995). }
\REF\dabholkar{A. Dabholkar and J. Harvey, Phys. Rev. Lett.
 {\bf 63}
(1989) 478; A. Dabholkar, G. Gibbons, J. Harvey, and F. Ruiz-Ruiz, 
Nucl. Phys. {\bf B340} (1990) 33.}
\REF\callan{C. Callan, J. Maldacena, and A. Peet,
hep-th/9510134 .}
\REF\waldram{A. Dabholkar, J. Gauntlett , J. Harvey, and D. Waldram,
hep-th/9511053.}
\REF\structure{F. Larsen and F. Wilczek,
hep-th/9511064 .}
\REF\cfthair{M. Cveti\v c and A. Tseytlin,
hep-th/9512031 , A. Tseytlin,
hep-th/9601177 .}
\REF\dyon{M. Cveti\v c and D. Youm,
Phys. Rev. {\bf D53} 584 (1996).}
\REF\canonicalrefs{R. Arnowitt, S. Deser, and C. W.  Misner,
{\it Gravitation: An introduction to Current Research} (L. Witten, Ed.),
Wiley, NY (1962); 
T. Regge and C. Teitelboim,
Ann. Phys. (N.Y.) {\bf 88} 286,
A. Hanson, T. Regge, and C. Teitelboim,
{\it Constrained Hamiltonian Systems},
Acc. Nat. Dei Lincei (Roma 1976).}
\REF\carlip{S. Carlip and C. Teitelboim,
Class. Quant. Grav. {\bf 12} 1699 (1995).}
\REF\deser{L. F. Abbott and S. Deser,  
Nucl. Phys. {\bf B195}, 76 (1982); Phys. Lett. {\bf B116}, 259 (1982). }
\REF\chandar{A. Balachandran, L. Chandar, and A. Momen,
hep-th/9512047 . }
\REF\teitelboim{C. Teitelboim,
Phys. Rev. {\bf D53} 2870 (1996) .}
\REF\holography{L. Susskind, 
J. Math. Phys. {\bf 36} 6377 (1995). }
\REF\toappear{F. Larsen and F. Wilczek, in preparation .}
\REF\cygeneral{M. Cveti\v c and D.  Youm, hep-th/9512127.}
\REF\exactcft{M. Cveti\v c and A. Tseytlin,
Phys. Lett. {\bf B366} 95 (1996).}
\REF\horowitz{ G. T. Horowitz and A. A. Tseytlin,
Phys. Rev. {\bf D51} 2896 (1995). }
\REF\chs{C. Callan, J. Harvey, and A. Strominger, 
Nucl. Phys. {\bf B367} (1991) 60.}
\REF\instanton{G. Gibbons and M. Perry, Proc. R. Soc. {\bf A358}
(1978) 467;~G. Gibbons and S. Hawking,
Comm. Math. Phys. {\bf 66} (1979) 291.} 
\REF\dbrane{A. Strominger and C. Vafa, hep-th 9601029,
C. Callan and J. Maldacena, hep-th/9602043,
G. Horowitz and A. Strominger, hep-th/9602051,
J. Breckenridge, R. Myers, A. Peet, and C. Vafa, hep-th/9602043,
J. Breckenridge, D. Lowe, R. Myers, A. Peet, A. Strominger, and
C. Vafa, hep-th/9603078,
C. Johnson, R. Khuri, and R. Myers, hep-th/9603061,
J. Maldacena and A. Strominger, hep-th/9603060,
R. Dijkgraaf, E. Verlinde, H. Verlinde, hep-th/9603126 .}


\chapter{Classical Hair}

The appearance of $\hbar$ in the denominator of the Bekenstein-Hawking
formula 
$$
S_{\rm BH}= {A\over 4\hbar G_N}~,
\eqn\bek
$$
suggests that, in a microscopic accounting, 
the entropy should be  visible classically.  
Indeed, a similar appearance 
of $(\hbar )^{-1}$ is familiar in ordinary gas dynamics, where it
provides 
a measure in classical phase space.  
In semiclassical quantization, one works with solutions of the
classical equations, which are parametrized by classical phase space.
One passes from these solutions to quantum states by requiring
quantization conditions of the Bohr-Sommerfeld type, and the spacing
of levels is therefore set by $\hbar^{-1}$.  Classical structure would
also be most welcome for another, related, reason.  If black holes
settle down to a unique (structureless) intermediate
state, independent of how they formed, then it
becomes impossible in principle to reconstruct the past state from the
future state.  Such a situation is difficult to reconcile with the
unitary evolution of states one expects in quantum theory.  The
difficulty is especially acute if the hole subsequently evaporates,
because one then appears to have an overall 
non-unitary evolution involving  ordinary matter only.  If black holes
were sufficiently structured -- if they had sufficient `hair' -- then
these problems might be avoided.


Of course the obvious difficulty with this straightforward, attractive way to
address these major issues in the quantum theory of black holes is the
famous meta-theorem that black holes have no hair: that is, that there
is a unique stationary classical solution for specified values of the 
conserved quantities (mass, angular momentum, and charges)
at infinity.  This theorem has been rigorously proved for
Einstein-Maxwell theory, and for small perturbations a fairly general
argument can be made [\wald ].  
It corresponds to the physical intuition that
gravitational collapse rapidly carries all the participating matter 
through the horizon, leaving behind only those traces that correspond
to surface integrals at infinity (e.g. charges, according to Gauss' law)
[\nohair ].  
The arguments used to establish the no-hair theorems are not entirely
general, and some isolated counterexamples are known in spontaneously
broken gauge theories [\hairrefs ].  
There are also interesting possibilities for structure  
of an essentially quantum-mechanical nature (quantum hair) in theories
with discrete gauge symmetries [\qhair ].   While these phenomena do serve to
emphasize that the no hair meta-theorem can fail in theories with
elaborate matter content, they do not appear to come close to providing the
massive degeneracy implicit in \bek .



Some recent developments, however, put this question in a new
light [\sen --\cfthair ].  
In analyzing black hole solutions for low-energy field
theories suggested by superstring theory, it is both interesting and
technically simplest to focus on solutions that preserve some 
supersymmetry. The equations that ensure supersymmetry, and {\it
inter alia\/} guarantee the equations of motion are obeyed, are first
order equations, and their general solution may be found.  When this
is done, one discovers that the solution contains some freely specifiable
functions. Our Appendix is devoted to reviewing a specific
example, the Cveti\v{c}-Youm dyon. In this regular black hole
solution --- a generalization of the extremal Reissner--Nordstr\"{o}m
black hole --- we exhibit hair of the form
$$
f(x_9, t) ~=~ f (x_9 - t)
\eqn\fncts
$$
where $t$ is Schwarzschild time and $x_9$ parametrizes a compactified
dimension, specifically a circle of radius $R$.  More precisely, the
hair is a product of \fncts\ with a profile function, and also
involves a change in the metric.  $f$ can be 
expanded
$$
f (x_9 - t) ~=~ 
f_0 + \sum_1^\infty a_n\cos ({2\pi n \over R }(x_9 -t )) +
   b_n \sum_1^\infty \sin  ({2\pi n \over R }(x_9 - t)) ~.
\eqn\modes
$$
>From the point of
view of a four-dimensional observer who does not resolve the extra
dimension, only $f_0$ is accessible.  It is a quantity that
must be specified as part of
the macroscopic description of the hole.  The remaining modes reduce
to zero upon dimensional reduction, but from the perspective of the
full theory they represent a potentially large number of hidden
degrees of freedom, in principle discernable at infinity.  
We anticipate that the phenomenon of
higher-dimensional hair is quite general, once one has an
appropriate field content. Indeed, the macroscopic fields at infinity
do not specify the dependence on compactified coordinates. Each of these
microscopic configurations generates a regular solution throughout
space-time exterior to the hole
by virtue of the no--hair theorem in the higher dimensional space.


Our main goal in this paper is to derive an effective theory
governing 
this sort of higher-dimensional hair, and to set up the
machinery for counting it (so as eventually to compare with \bek ).
An important point of the analysis is the derivation, from imposition
of the
requirement of vanishing source at the horizon, of a matching
condition which forces the existence of non-trivial hair.  In the
companion paper [\toappear ], 
we will apply this machinery to the Cveti\v{c}-Youm 
dyon.


It is appropriate now to outline the logic of the remainder of this
paper.  

We wish to exhibit and eventually count the effective low-energy
degrees of freedom for specified quantum numbers in a complicated
field theory containing many other real and fictitious (gauge) degrees
of freedom.  To do this it seems inevitable that we must cast the
theory of interest -- containing gravitational, dilaton, gauge, and
antisymmetric tensor fields -- in canonical form.  Issues involving the
choice of surface terms are particularly important for us.  We find
that general covariance and duality, 
important properties of the bulk theory, lead us to a
definite choice.  As we formulate a dynamical theory based on
fields entirely outside the horizon the horizon appears formally
as a spatial (null) boundary.  Our complete specification of surface terms
guides us toward 
appropriate boundary conditions at such a  boundary.  

Next we 
construct a reduced Hamiltonian, that depends only on the parameters
of the classical solution -- the hair variables.  There is a general
procedure for extracting such a reduced Hamiltonian, which we will
discuss.
In our context, the result
assumes a very characteristic form.  The bulk Hamiltonian vanishes as
a consequence of
reparametrization invariance, which imposes constraints.  The reduced
Hamiltonian therefore consists entirely of surface terms and, since
the surface terms at infinity are of a very simple form, the dynamics is
in this concrete sense localized to the horizon.  However in solving
the constraints one finds that some of the surface variables, when
expressed in terms of the hair variables, involve integral
expressions (over a fixed spatial profile), 
so that in another sense the dynamics extends outside.
The form of the outer fields is quite restricted.  After integrating it out 
one reaches, in the case of the
Cveti\v{c}-Youm dyon, an effective string theory for the classical hair.

As a by-product of our development we obtain in a canonical fashion
formal expressions for the entropy that were previously deduced using
Euclidean methods, which appears to us to be a conceptual advantage.




\chapter{Canonical Formalism for Dilaton Gravity}

In this section, we consider canonical treatment
of the bosonic part of heterotic string 
theory compactified to $D$ dimensions on a torus with constant moduli. 
The Lagrangian density is 
$$
16\pi G_N{\cal L} = \sqrt{-G}e^{-2\Phi}[
R^{(D)} + 4(\nabla\Phi)^2 - {1\over 12}H^2 - {1\over 4}\alpha^\prime F^{(i)2} 
] + \partial_I V^I
\eqn\lag
$$
where\foot{Index conventions are $I,J,\cdots = 0,\cdots,D-1$ and
$\alpha,\beta,\cdots =1,\cdots,D-1$ for the spacetime indices
and $i=1,\cdots,16$ for the internal ones. 
In the following $\alpha^\prime=1$.}
$$\eqalign{
H_{IJK}&=(\partial_I B_{JK}-{1\over 2}\alpha^\prime A^{(i)}_I F^{(i)}_{JK})
+({\rm cyclic~permutations})\cr
F^{(i)}_{IK}&=\partial_I A^{(i)}_K - \partial_K A^{(i)}_I \cr}
\eqn\HF
$$
The total derivative 
$$
\partial_I V^I = 4{D-1\over D-2} \partial_I 
(\sqrt{-G}e^{-2\Phi} \partial^I \Phi )
\eqn\surfacelag
$$
will be discussed in the following chapter. 

The linchpin of a canonical formalism is 
division of the metric into spatial and temporal parts.
We consider the explicit form 
$$
dS^2 = -(Ndt)^2+g_{\alpha\beta}(dx^\alpha+N^\alpha dt)(dx^\beta+N^\beta dt)
\eqn\lapse
$$
General covariance implies that there is great 
arbitrariness in the choice of lapse and shift functions $N$ 
and $N^\alpha$. 

One finds the field momenta 
by varying $16\pi G_N L$ with 
respect to $\partial_t g_{\alpha\beta}$, $\partial_t \Phi$, 
$\partial_t A_\alpha$, and $\partial_t B_{\alpha\beta}$. 
By expanding the Lagrangian \lag~, assuming the metric \lapse~, and
performing the variations we find
$$\eqalign{
\Pi^{\alpha\beta}&=g^{1\over 2}e^{-2\Phi}[g^{\alpha\beta}{\rm Tr}K
-K^{\alpha\beta}] +{2\over N}g^{\alpha\beta}g^{1\over 2}e^{-2\Phi}
(\partial_t \Phi-N^\gamma
\partial_\gamma \Phi )\cr
\Pi^\Phi &= -{8\over N}g^{1\over 2}e^{-2\Phi}[\partial_t\Phi-N^\alpha
\partial_\alpha \Phi ] -4g^{1\over 2}e^{-2\Phi} {\rm Tr}K \cr
\Pi^{(A)}_\alpha &\equiv {\cal E}_\alpha=
{1\over N}g^{1\over 2}e^{-2\Phi}[(F_{t\alpha}-N^\beta F_{\beta\alpha})
+{1\over 2}(H_{t\alpha\beta}-N^\gamma
H_{\gamma\alpha\beta})A^\beta ] \cr
\Pi^{(B)}_{\alpha\beta}&\equiv {\cal E}_{\alpha\beta} =
{1\over 2N}g^{1\over 2}e^{-2\Phi}[H_{t\alpha\beta}-N^\gamma
H_{\gamma\alpha\beta}] \cr}
\eqn\momenta
$$
where the extrinsic curvature is
$$
K_{\alpha\beta}={1\over 2N}[N_{\alpha |\beta}+N_{\alpha |\beta}
-\partial_t g_{\alpha\beta}]
\eqn\extcurv
$$
The stroke denotes covariant derivative with respect to the spatial
metric. The detailed calculations leading to \momenta~ and to the
equations below are rather involved. 
A good strategy for these calculations,
and some useful identities, are presented in [\canonicalrefs ].

The Hamiltonian is defined by the Legendre transform
$$\eqalign{
16\pi G_N H&=\Pi^{\alpha\beta}\partial_t g_{\alpha\beta}
+\Pi^\Phi \partial_t \Phi + {\cal E}^\alpha \partial_t A_\alpha
+{\cal E}^{\alpha\beta} \partial_t B_{\alpha\beta}-16\pi G_N{\cal L} \cr
&=N{\cal H}+N_\alpha {\cal H}^{\alpha}
-A_t C - B_{t\alpha}C^{\alpha} + 16\pi G_N {\tilde H} 
\cr}
\eqn\formalh
$$
Let us explain the general form indicated in the last line. 
The gauge fields $A_t$ and $B_{t\alpha}$ 
have no associated kinetic terms, so they act as Lagrange multipliers. 
The corresponding constraint equations generalize Gauss' law. 
They are $C=C^\beta=0$ where
$$\eqalign{
C&=\partial_\alpha {\cal E}^\alpha -{1\over 2}{\cal E}^{\alpha\beta}
F_{\alpha\beta} \cr
C^\beta &=2\partial_\alpha {\cal E}^{\alpha\beta}
\cr}  \eqn\constr
$$
The lapse and shift functions $N$ and $N^{\alpha}$ are metric
analogues of these non--dynamical variables. They enforce the
constraints ${\cal H}={\cal H}_{\alpha}=0$ where
$$\eqalign{
{\cal H}&=  g^{-{1\over 2}}e^{2\Phi}
[{\rm Tr}~\Pi^2 +{1\over 2}\Pi^\Phi {\rm Tr}~\Pi +{D-2\over 16}\Pi^2_{\Phi}]
-g^{1\over 2}e^{-2\Phi} R^{(D-1)} \cr
&
+4g^{1\over 2}e^{-2\Phi}g^{\alpha\beta}\partial_\alpha\Phi
\partial_\beta\Phi - 4e^{-2\Phi}\partial_\beta (g^{1\over 2}
\partial^\beta \Phi ) \cr
&+{1\over 2}~g^{-{1\over 2}}e^{2\Phi}
g_{\alpha\beta}({\cal E}^\alpha + {\cal E}^{\alpha\gamma}A_\gamma)
({\cal E}^\beta +{\cal E}^{\beta\delta}A_\delta)
+{1\over 4}~g^{1\over 2}e^{-2\Phi}F_{\alpha\beta}~F^{\alpha\beta}+ \cr
&+g^{-{1\over 2}}e^{2\Phi}{\cal E}_{\alpha\beta}{\cal E}^{\alpha\beta}+
{1\over 12}g^{1\over 2}e^{-2\Phi}H_{\alpha\beta\gamma}H^{\alpha\beta\gamma}
\cr}
\eqn\hamilton
$$
and
$$
{\cal H}_{\alpha}= -2{\Pi}^{~\beta}_{\alpha~~|\beta} +\Pi^\Phi 
\partial_\alpha \Phi+
F_{\alpha\beta}({\cal E}^{\beta} + {\cal E}^{\beta\gamma}A_\gamma )
+H_{\alpha}^{~\beta\gamma}{\cal E}_{\beta\gamma}
\eqn\supermom
$$
In each case the constraint is identified by varying with
respect to the appropriate Lagrange multiplier. 

Alternatively, the constraints can be viewed as a generators of
continuous symmetries in the space of all field configurations.
The symmetries are implemented by insisting that the generators
act trivially. This explicit identification of symmetry generators
is an important aspect of the canonical formalism.  
The Lagrangian formalism, in contrast, does not distinguish
the constraints from other equations of motion.

The final term in the Hamiltonian, $\tilde{H}$, differs qualitatively 
from all the preceding
constraint terms. It is the total derivative
$$\eqalign{
16\pi G_N {\tilde H}&={2\over D-2}\partial_t {\rm Tr}~\Pi + 
2\partial_\alpha (
\Pi^{\alpha\beta}N_\beta - {1\over D-2}{\rm Tr}~\Pi ~N^\alpha ) + \cr
&+2\partial_\alpha [Ng^{1\over 2}e^{-2\Phi}g^{\alpha\beta}
({1\over N}\partial_\beta N -{2\over D-2}\partial_\beta \Phi)]
+\partial_\alpha (A_t {\cal E}^\alpha )\cr  &+2\partial_\alpha
(B_{t\beta}{\cal E}^{\alpha\beta})
\cr}
\eqn\surfaceham
$$
This term includes both
contributions from the explicit derivative term 
in the original Lagrangian and terms that follow
from integrations by parts. It plays a crucial role
in our considerations. 
To avoid any ambiguity let us emphasize  that {\it all\/} 
terms have been retained.

The relations \momenta ~are recovered by
varying the Hamiltonian with respect to the momenta and 
simplifying the equations. This serves
as an important check on the algebra. The remaining
equations of motion are found by variation with respect to the fields.
The resulting expressions are very lengthy and will not be displayed
here (nor used below).

\chapter{Duality and Surface Terms}

The total derivative $\partial_I V^I$ in the Lagrangian \lag ~is 
not determined by classical macroscopic physics,
because it does not affect the equations of motion.
It must, however, be specified to define the action off shell, and can
play a role in specifying the quantum theory.  It is 
ordinarily fixed by choosing appropriate boundary conditions
and requiring that the complete Hamiltonian is stationary under 
all variations  which satisfy them [\carlip ,
\teitelboim ]. 
Here we proceed quite differently: we resolve the ambiguity by 
demanding invariance under a symmetry: duality.
Under duality the metric and dilaton transform as
$$
G_{IJ}\rightarrow e^{-2\gamma\Phi} G_{IJ}~~~~~~~\Phi\rightarrow -\Phi
\eqn\lagdual
$$
where $\gamma={4\over D-2}$. Under this transformation
$$
\sqrt{-G}e^{-2\Phi}R^{(D)} \rightarrow \sqrt{-G}e^{-2\Phi}R^{(D)}
+8{D-1\over D-2} \partial_I (\sqrt{-G}e^{-2\Phi} \partial^I \Phi )~.
\eqn\rtransdual
$$
The total derivative \surfacelag ~was chosen such that it exactly
cancels the inhomogeneous term in \rtransdual ~, 
so that the graviton-dilaton portion of the Lagrangian is invariant.
The corresponding condition in Einstein frame is the absence of an
explicit surface term.  In this manner
the Lagrangian is specified uniquely by general 
covariance and duality, even off shell.

In the canonical formalism time and space are treated differently, and 
so general covariance is no longer manifest. Duality, on the other
hand, can be expressed explicitly in terms of canonical variables. 
Indeed, the transformation
$$\eqalign{
g_{\alpha\beta}&\rightarrow e^{-2\gamma\Phi} g_{\alpha\beta} \cr
\Pi^{\alpha\beta}&\rightarrow e^{2\gamma\Phi} \Pi^{\alpha\beta} \cr
\Phi &\rightarrow -\Phi \cr
\Pi^\Phi &\rightarrow -\Pi^\Phi -2\gamma {\rm Tr} \Pi \cr
N&\rightarrow e^{-\gamma\Phi}N \cr
N^\alpha &\rightarrow N^\alpha \cr}
\eqn\dualh
$$
leaves $N{\cal H}$, ${\cal H}_\alpha$, and ${\tilde H}$
separately invariant\foot{We have not specified the
transformation of the matter fields.  In all cases known to us, useful
duality transformations do not 
relate the graviton-dilaton to other sectors.}. 
It also preserves the Poisson brackets,
so that it is a canonical transformation. 
As one consequence, the measure in
classical phase space is invariant under
duality.

Having determined the surface terms by duality it is meaningful
to reverse the ordinary procedure, and
use the variational principle to discover the appropriate 
boundary conditions.
The equations of motion follow from the Hamiltonian 
after variation and integration by parts. 
The surface terms thus generated must cancel
the variation of the explicit surface terms in order that
the bulk equations truly represent the conditions for stationarity.

First consider the surface terms at the horizon.  Let us 
choose an
adapted coordinate system where the metric is of the form
$$
dS^{2} = -(Ndt)^2 + d\rho^2 + \gamma_{mn} (dx^m + N^m dt)(dx^n + N^n dt)~.
\eqn\bhmetric
$$
Here $\rho$ is the normal coordinate close to the
horizon and the $x^m$ are transverse coordinates.
All black holes can be written in this form.
For this metric the spatial curvature is
$$
R^{(D-1)}= -2 \gamma^{-{1\over 2}} \partial_\rho^2 \gamma^{1\over 2}
+{1\over 4}[(\gamma^{mn}\partial_\rho \gamma_{mn})^2
-\gamma^{km}\partial_\rho \gamma_{mn}\gamma^{nl}\partial_\rho \gamma_{lk}]
+\cdots
\eqn\spaccurv
$$
where $\gamma$ denotes the determinant of the transverse metric $\gamma_{mn}$
and the omitted terms contain no $\rho$-derivatives.

For simplicity, let us first consider the situation when $\Phi=0$. 
Then the variation of the Hamiltonian is
$$
\delta H = \partial_{\rho} [2N\partial_{\rho} 
\delta\gamma^{1\over 2}-{1\over 2}N\gamma^{1\over 2}
(\gamma^{kl}\partial_\rho \gamma_{kl}\gamma^{mn}-\gamma^{mk}
\partial_\rho \gamma_{kl}\gamma^{ln})\delta\gamma_{mn}]
+{\rm bulk~term}
\eqn\hamvar
$$
The bulk term gives the equation of motion. The condition
for the boundary term to vanish is complicated. This feature reflects
that dynamics at an arbitrarily specified surface must be
highly non--trivial. We are however -- not coincidentally --
interested in the exceptional case where $N=0$ on the surface, 
when things greatly simplify. $N=0$ 
is a very strong condition in Minkowski space. 
It implies that $t$ is null\foot{Up to reparametrizations. Shift functions
have been suppressed to simplify notation but should be retained
as a matter of principle.}. 

The boundary is generated by $t$, so when $t$ is null the boundary is,
according to a  theorem of Penrose, a
future event horizon. 
Moreover, Hawking's theorem states that on the solution
$$
{\partial\gamma^{1\over 2}\over\partial\lambda} \geq 0
\eqn\secondlaw
$$
where $\lambda = Nt$ is the affine parameter.  (Sufficient positive
energy conditions are satisfied in the classical field theory we are 
considering.)
For reversible processes this is an equality
that amounts to the condition
$$
\Pi^\rho_\rho =0
\eqn\pirhorho
$$
by the definition \momenta . 
This requirement is a consequence of the causal
structure of spacetime and of general positivity requirements,
rather than a separate dynamical principle. 

In summary,
the boundary conditions $N=0$ gives an off--shell definition of 
a black hole in Minkowski space. There is an additional condition
$\Pi^\rho_\rho=0$ for reversible processes.  The boundary 
conditions on the metric $\gamma_{mn}$ were free, so there was no need 
to take $\partial_\rho\gamma^{1\over 2}=0$. This is the condition for
an apparent horizon at the boundary. It may be satisfied dynamically
for a specific solution, 
but in general it is not.  We find it  reasonable 
that the boundary must be a future event horizon for
a good Cauchy problem to be posed, while the existence of an 
apparent horizon is a dynamical question.  The boundary conditions we
have obtained reflect this.

The remaining gravitational equations of motion are found by varying
the momenta. The surface terms in this case lead to the requirement 
that the shifts
$N^\alpha$ at the horizon must be kept fixed as the momenta are varied. 
On the other hand 
it is crucial for our later purposes to note that no restriction is placed on 
their value. 

Finally, a general value of the dilaton $\Phi$ should 
be restored. 
This can be accomplished by transforming to the Einstein frame
$$\eqalign{
g_{\alpha\beta}^E &=e^{-{2\over D-2}\Phi}g_{\alpha\beta} \cr
N^E &= e^{-{1\over D-2}\Phi}N \cr}
\eqn\einsframe
$$
In this frame the preceding 
equations are valid without amendment. The variational principle for
the dilaton can be derived at fixed Einstein metric. The boundary 
condition becomes $\Phi$ fixed on the horizon. 
This is curious because a
fixed value of $\Phi$ is not duality invariant unless it is $\Phi =0$ . 
Therefore duality must be violated either at the horizon or,
preferably from the present point of view, at infinity.

Having discussed the surface terms at the horizon, we now
turn our attention to the 
surface terms at infinity. Again, there is a unique term that follows from
general covariance and duality. Unfortunately the ensuing boundary 
conditions are so restrictive that only flat space is a solution!
This is analogous to a situation that arises in gauge theories: since
charged states are gauge non-singlets, to specify a charged state --
which is a very physical concept -- one must formally violate gauge 
invariance [\deser ].
Concretely, it is
necessary to supplement the surface terms at infinity with
non-invariant terms added by hand. For example, the term
$$
H_{\infty}= M_{\rm ADM}
\eqn\hinf
$$
is appropriate to allow solutions with ADM mass $M_{\rm ADM}$.
The explicit expression for the ADM Hamiltonian can be found in 
the literature [\canonicalrefs ]. The central issue is readily 
illustrated by considering a free gauge field: 
the boundary term that follows from integration by parts prior to
variation, and the
one that follows from variation, differ by a factor two. For gravity
the details are more complicated but again
the ADM Hamiltonan \hinf\ turns out to be 
twice the value that follows from the unamended surface Hamiltonian.
The term at infinity must always be chosen to be finite, and 
inequivalent choices are related by Legendre transforms
of the generating functional. In the context of black hole thermodynamics
it is possible to transform from the microcanonical to the canonical
ensemble, for example. It is natural to retain this freedom by
not committing to a specific term at infinity. 

The breaking of general covariance by the boundary at infinity is a 
well-known subtlety, not unrelated to Mach's principle [\deser ]. 
Poincare invariance is a subgroup of the full symmetry that can be 
restored [\canonicalrefs ]. For our work the most important point is that
the term at infinity can be chosen to be independent of our
classical hair, so that general covariance in the compactified 
space remains unbroken.
 
Since general covariance is broken by the boundary at infinity it
is a logical possibility that it is also broken by the boundary 
at the horizon, as has been explored by Teitelboim
[\teitelboim ]\foot{see also [\chandar ] for a recent alternative
approach.}.  We are proposing the principle that at a more
microscopic level this does not happen.  We find it an attractive hypothesis
-- and, as will appear, one with very non--trivial consequences -- 
that sources 
of macroscopic charges can be  generally covariant in this sense.

\chapter{The Reduced Hamiltonian}

The full Hamiltonian contains the equations 
of motion for all physical degrees
of freedom, as well as many redundant variables. 
In discussing the low-energy dynamics we propose for black holes 
we are primarily interested in a very small subset, 
namely the variables describing the hair. The reduced problem
is still rather complicated, because we must take into account how
the other degrees of freedom affect the hair.
Hamiltonian reduction is
the appropriate formalism for this problem [\canonicalrefs ]. 

Abstractly, Hamiltonian reduction  works
as follows.  Denote the variables pertaining to the hair 
$(\pi_a , \phi_a )$, and  the remaining variables which
describe the background
by $(\pi_A , \phi_A )$ .  
The equations of motion for the hair are the canonical equations
\foot{$\delta$ denotes the variational derivative
so here the symbol $H$ denotes the spatial integral of the density given
in the previous chapters. This should not cause any confusion.} 
$$
{\dot\phi_a} = {\delta H (\phi_a, \pi_a, \phi_A, \pi_A )\over 
\delta \pi_a}~~;~~~~
{\dot\pi_a} = -{\delta H (\phi_a, \pi_a, \phi_A, \pi_A )\over 
\delta \phi_a} \eqn\caneqs
$$
In these equations the background variables  $(\pi_A , \phi_A )$ should
be treated as independent variables and therefore kept fixed.

There is an alternative derivation of the same equations.
Consider 
$\phi_A$ to be given static functions
and let the canonically conjugate variables $\pi_A $ be solutions of
the constraints given these $\phi_A$. $\pi_A $  will be a function of
the prescribed functions $\phi_A$ and of the hair variables.
One has
$$
{\delta H (\phi_a, \pi_a , \phi_A, \pi_A )\over 
\delta \pi_a}
={\delta H (\phi_a, \pi_a , \phi_A, \pi_A )\over 
\delta \pi_a}_{|\pi_A}+
{\delta H (\phi_a, \pi_a , \phi_A, \pi_A )
\over \delta \pi_A} {\delta \pi_A \over \delta \pi_a}
\eqn\cander
$$
In the first term $\pi_A$ is kept fixed, just as it was in \caneqs .
In the second term it is varied; however the canonical 
equations for the background fields equate this term to the time
derivatives of $\phi_A$, which were assumed to vanish. 
Thus the alternative
way of performing the variation gives the same result as the
original one. It is easy to show that 
this feature is shared by the variations with respect to the
fields $\phi_a $.  

In this way, we have identified a Hamiltonian that
generates the correct equations of motion. 
It contains only the reduced (i.e. for us, hair) variables as
dynamical degrees of freedom, and is known as the reduced Hamiltonian.

$\pi_A$ is chosen to be a solution of the constraints off-shell 
even when the 
hair coordinates and momenta do not satisfy the equation of motion. 
Since the constraints vanish off-shell, 
the reduced Hamiltonian like the full Hamiltonian simplifies to 
a the total derivative term
$
H_{\rm reduced}=\tilde{H}
$ 
where $\tilde{H}$ is the total derivative \surfaceham . 
This enormous reduction is a general feature of systems with general
covariance.

The reduced Hamiltonian is a total derivative; thus formally
it can be evaluated in terms of 
quantities living  on the boundaries of spacetime. 
However the constraints are differential equations, 
so their integrals --
and thereby  the reduced Hamiltonian --  can nevertheless contain 
information about all of space. Concretely in the present 
context, this
means that the reduced Hamiltonian can include information about the radial
profile of the hair. 

However, because of the special form \modes\ of 
the hair, if it is specified at any radius,
it will be uniquely determined throughout spacetime. 
This feature embodies, in our higher-dimensional context,
the physics of the conventional no hair theorem. 
Thus the profile functions do not represent 
any independent dynamical variables and it is natural
to regard the hair {\it on the horizon\/} 
as the proper reduced variables. 

The reduced Hamiltonian can still be a non--local function of the transverse 
directions on the horizon.  We expect that this 
non--locality is tightly restricted
by the special properties of global horizons, 
but this point needs further investigation. 

The effective surface Hamiltonian represents the low energy 
dynamics of spacetime. There are equivalent representations
that project on to any reasonable surface surrounding the black hole.
This appears to implement something like Susskind's 
holographic principle in a natural 
manner [\holography ].  Intricate interplay between constraints
and apparent non-locality arises  appears to be an inevitable
consequence of reduced dynamics in a generally covariant system.

Before concluding this section we should mention 
the non--dynamical fields (e.g. the lapse and shift)
that appear prominently in the metric, but appear to be quite 
secondary in the Hamiltonian framework.  
They do not enter the constraints explicitly, so they 
should not be specified in solving for the momenta.
The equations of motion that follow from variation with respect to 
the background momenta were used crucially in the derivation
of the reduced Hamiltonian. They require that the 
Lagrange multipliers are fixed during the variation, but left their
values undetermined. The residual 
freedoms in the surface Lagrange multipliers 
implement symmetries of the effective surface theory.

 
\chapter{Black Hole Statistical Mechanics}

After these preparations, we are now prepared to discuss
some specific features of the
surface theory.  For ease of reference 
we repeat the Hamiltonian \surfaceham~ 
$$\eqalign{
16\pi G_N{\tilde H}&={2\over D-2}\partial_t {\rm Tr}~\Pi + 2\partial_\alpha 
(\Pi^{\alpha\beta}N_\beta- {1\over D-2}{\rm Tr}~\Pi N^\alpha ) + \cr
&+2\partial_\alpha [Ng^{1\over 2}e^{-2\Phi}g^{\alpha\beta}
({1\over N}\partial_\beta N -{2\over D-2}\partial_\beta \Phi)]
+\partial_\alpha (A_t {\cal E}^\alpha ) +2\partial_\alpha
(B_{t\beta}{\cal E}^{\alpha\beta})
\cr}
\eqn\surfaceha
$$
Each
term should be considered, in the reduced description, as a function of the hair and
the macroscopic quantum numbers that describe the background.
The reduced Hamiltonian is the spatial integral of this
expression.

The matter terms 
will not be considered here.  For the rest,
we find in
the coordinates \bhmetric~ 
$$
16\pi G_N (H_{\rm reduced}-M_{\rm ADM}) = -2 \int_{\rho=0} d^{D-2}x ~[
\Pi^\rho_\alpha N^\alpha + \gamma_E^{1\over 2} \kappa ]~.
\eqn\formalhred
$$
Here
$$
\kappa = \partial_\rho N^E
\eqn\surfaceacc
$$
is the surface acceleration in the Einstein frame, and we chose
the surface term \hinf~ at infinity for definiteness. 
According to the variational principle discussed in 
the previous chapter, $\kappa$
and $N^\alpha$ must be considered constants when varying other fields.

Each quantity in \formalhred ~is separately invariant
under duality.

The lapse functions $N^\alpha$ are arbitrary and do not depend on the hair
variables. This reflects reparametrization symmetry of 
the surface theory, as we have discussed. 
Variation with respect to $N^\alpha$ gives
$$
\Pi^\rho_\alpha = 0~~;~~~\alpha \neq \rho
\eqn\matching
$$
The momenta $\Pi^\rho_\alpha$ serve as
generators of coordinate transformations in the surface
theory, so these constraints are analogous to the
constraints in the bulk theory.  These equations take the form 
of {\it matching 
conditions\/} relating properties of the background to properties of 
the hair. For black hole solutions that derive from the fundamental
string it agrees with the condition that can be derived by
matching onto a source [\waldram ], or from cosmic censorship
[\callan ].

In the theory without hair the momenta $\Pi^\rho_\alpha$ with
$\alpha=4,\cdots,D-1$ would be interpreted as a manifestation
at the horizon of Kaluza--Klein charge at infinity. The hair modifies 
this, and our matching conditions relate the amplitudes
of the hair to the background charges in such a way that the combined 
momenta vanish; this implements the `no-source' boundary condition we
discussed earlier.

The matching conditions have important consequences. 
First, they further reduce the number of independent
variables.  This is because they embody the requirement 
that hair variables that
differ only by reparametrization are physically equivalent.
Second and more profoundly, they delegitimize
the original bald black hole, which
does not satisfy the constraint. 
In order that all constraints be satisfied, 
there {\it must\/} be hair.

Analogously, for a rotating black hole without
hair the variables $\Pi^\rho_\theta$ 
and $\Pi^\rho_\phi$ are non--zero at the horizon, so the matching 
conditions \matching~ seem to rule out angular momentum. 
Actually what is excluded 
is pure rotation, which would violate general covariance at the boundary.
The rotation should be matched with appropriate hair, so that the combined 
system is generally covariant. 
General
covariance on the horizon appears to be so strong that, 
for black holes without hair, it
excludes interesting physics and must be relaxed. In a complete
theory including hair it appears it can be maintained, however. This provides
an appealing {\it raison d`\^{e}tre} for the hair.

Now consider the second term in \formalhred. In this paragraph 
we shall proceed 
schizophrenically, ignoring the presence of hair,
in
order to make contact with previous understanding.  We have already 
met the surface acceleration 
$\kappa$ in \surfaceacc .  It is duality invariant and
constant on the horizon, which suggests its independent physical
significance.  
Let us introduce
the formal temperature $\Theta = {\kappa\over 2\pi} $, which
is of course
also  constant
along the horizon. Then the integral over angular variables gives
$$
H_{\rm reduced}-M_{\rm ADM} = -{A_E \over 4G_N }~\Theta~~;~~~A_E\equiv
\int_{\rho=0} d^{D-2}x ~\gamma_E^{1\over 2}
\eqn\hrdebh
$$
In this expression the area can be considered either the four dimensional
area of the black hole or the higher dimensional one. They differ by a
factor of
the volume of the internal space, which should be absorbed in $G_N$
under dimensional reduction. The reduced Hamiltonian is the
generator of equations of motion with the temperature $\Theta$
kept fixed {\it i.e.} the free energy $F=E-\Theta S$. 
The appropriate surface term at infinity, which keeps the temperature fixed, 
is the ADM mass that was already included above. Hence we
recover the celebrated Bekenstein--Hawking formula
$$
S={A_E \over 4G_N }
\eqn\bekhw
$$

The present derivation was made at a given spatial section. Therefore
no Euclideanization was necessary, and no conical angle was introduced.
There were also no infinite terms in need of regularization
and subtraction. These are attractive features of this derivation
but several heuristic elements remain: $\Theta$ as measured using
surface acceleration was used as a macroscopic variable without
formal justification. Its normalization was not fixed by any
consideration internal to the calculation; this can almost certainly
be remedied by inserting an appropriate thermometer, {\it i.e}. by
carefully quantizing the modes of a model field and detector at infinity.  

Most importantly, the derivation had no explicit
reference to the microscopic degrees of freedom that gave rise
to the entropy.  The classical hair is proposed to remedy this. In fact,
the Bekenstein--Hawking formula has a somewhat unusual interpretation
in the present framework. The area is dynamical and {\it a priori} it
has no independent physical meaning. It is a function of the
hair variables that only reduces to the area when the hair is
disregarded. 
Indeed, in the Cveti\v{c}-Youm example the nine-dimensional area vanishes
at the origin by virtue of the matching condition (but the
three-dimensional spatial metric is unaffected.)
We suspect that, when
fully spelled out,  \bekhw\ will be appear directly as
a phase space integral over the microscopic degrees of freedom.
At present we do not know how to do this in generality {\it i.e.}
without considering the explicit form of the constraint
$\Pi^\rho_\alpha =0$.  We shall carry this program out further in
[\toappear ].

Finally, consider the terms that are total
derivatives with respect to compactified 
coordinates.  Upon integrating the Hamiltonian over a spatial section
such terms vanish when the fields are periodic functions of
the coordinates.  
But potentials, as opposed to field strengths, 
need not be periodic in general, 
and the total derivative terms
can encode topological information. 
Concretely, assume that there is a string's worth of hair, as 
occurs for the example discussed in the Appendix.  In fundamental
string theory the 
worldsheet coordinate can map into spacetime with a non--trivial 
winding number. 
In the present spacetime approach there is no world sheet and no
fundamental string.  Nevertheless there may be winding. 
This would manifest itself through non-vanishing contributions from the
total derivative terms.  It is intriguing that the same quantity
${\rm Tr~}\Pi$ ($\propto P_1 P_2 $ for the Cveti\v{c}-Youm dyon) 
sets the dynamical scale both for the compactified dimension $x_9$ and
time
$t$ -- or eventually, inverse temperature.

\chapter{Comments}

We have already made several interpretational remarks in the
text; we wish only to add  two brief comments.

Although most of the formalism in the main text is valid more
generally, our working example involves extremal black holes.
It is quite plausible that the physics underlying formulation of
appropriate matching conditions and counting of states is more
complicated for non-extremal holes; and specifically, that their
non-zero temperature plays a crucial role.  Indeed, the spirit of the
matching condition is that the specified charges at infinity require,
for a solution which is in a strong sense source-free at the horizon, 
a definite non-zero amount of hair outside.  It has long seemed a
striking coincidence that the energy of the classic 
extremal Reissner-Nordstr\"{o}m
hole can be regarded as being entirely in electrostatic fields
outside the horizon, according
to $M = Q^2/R$ for $Q=R=M$.  Thus there is at least one rough sense in
which 
there's `nothing inside', as we have proposed in a somewhat 
different form.
For non-extremal holes the qualitative picture is much less clear;
one might speculate that the matching condition reflects the necessity not
only to make up the requisite charges but also to provide the
appropriate thermal excitations in a self-consistent manner.

Recently there have been some truly remarkable developments in the study of
black holes closely related to ones we consider here generally and to
the Cveti\v{c}-Youm dyon in particular [\dbrane ].  
(In its details most of the work has
focussed on five dimensional versions, for reasons that
are technical and presumably temporary.)  The main thrust of this work
is to use D-brane technology to count BPS saturated states with
certain quantum numbers in the weak-coupling limit, and then to argue
that it is valid to extrapolate this counting to strong coupling,
when the states become black holes.  The approach suggested here in
no way contradicts these developments, but attempts to deal more
directly with the space-time aspects of the problem.  Particularly
when the states under consideration are macroscopic black holes, a
classical or semi-classical approach ought to be
appropriate and convenient.  We have argued that
classical hair exists in abundance, and have emphasized
its potential for addressing the classic problems of microstate
counting and information storage.

{\bf Acknowledgements}
We are grateful to C. Teitelboim for important discussions.

\endpage

\refout 
 
\endpage


\appendix

\section{The Cveti\v{c}--Youm dyon}
In this appendix we explicitly construct hair on 
the Cveti\v{c}--Youm dyon [\dyon ].
This is a spherically symmetric four dimensional black hole 
solution to low energy heterotic string theory that has $N=1$
supersymmetry. It is almost -- but not quite -- the most
general solution with these properties [\cygeneral ].
It can be considered an exact conformal field theory [\exactcft ]. 
The black hole exhibits remarkable features which strongly suggest
that all its entropy can be accounted for by the mechanism pursued
here [\structure ]. 

The black hole is parametrized by 4 independent charges.
The line element is, explicitly,
$$\eqalign{
dS^2 &= F du ( dv + K du ) +G_{ij}dx^i dx^j \cr 
G_{ij}dx^i dx^j  &= f [ k 
( dx^{(4)} + {\bf P^{(1)}}(1-\cos\theta)d\phi)^2
+ k^{-1}( dr^2+ r^2 (d\theta^2 +\sin\theta^2 d\phi^2 ))] \cr
&+ \sum_{i,j=5}^{8} \delta_{ij}dx^i dx^j \cr}
\eqn\stringmetric
$$
in string metric. Here $u=x^{(9)}-t$, $v=x^{(9)}+t$, and
$$
F^{-1}=1+ {{\bf Q^{(2)}}\over r}~~;~~~K= {{\bf Q^{(1)}}\over r}~~;~~~
f=1+ {{\bf P^{(2)}}\over r}~~;~~~k^{-1}=1+{ {\bf P^{(1)}}\over r}
\eqn\fkfk
$$
The other nonvanishing fields are
$$
B_{uv}=F~~;~~~B_{\phi 4}={\bf P^{(2)}}(1-\cos\theta)~~;~~~e^{\Phi}=Ff
\eqn\bbdil
$$
We will also use the notation $\Phi_\parallel={\rm ln}F$ and
$\Phi_\perp={\rm ln}f$.
Each compact dimension gives rise to two $U(1)$ gauge fields: one
from the metric, and one from the antisymmetric tensor field. In the
solution above, the $U(1)$'s from the 9th dimension are assigned
electric charges, those from the 4th dimension are assigned magnetic ones,
and the remaining sectors are neutral. The standard extremal
Reissner--Nordstr\"{o}m black hole is realized as the special case
${\bf Q^{(1)}}={{\bf Q^{(2)}}=\bf P^{(1)}}={\bf P^{(2)}}$, when the
dilaton decouples.  It appears, of course, in its Kaluza-Klein form.

The Cveti\v{c}--Youm dyon combines many of the known 
solutions of low energy string theory:
$F\neq 1$ is characteristic of the fundamental string [\dabholkar ], 
$K\neq 0$ of the plane wave [\horowitz ], 
and the combination of the two of the charged fundamental string 
[\dabholkar, \callan, \waldram ]. 
$f\neq 1$ is the symmetric five--brane [\chs ]
and $k\neq 1$ is the self--dual taub-NUT gravitational instanton [\instanton ].
It is remarkable that all these solutions can coexist as they do here.

By generalizing the calculation in the appendix of [\horowitz ],
one can  show that the equations of motion for the {\it ansatz } 
\stringmetric ~reduce to a number of Laplace equations with the 
solutions \fkfk . For example, one equation is
$$
{1\over ({r+{\bf P^{(1)}}})({r+{\bf P^{(2)}}})}{\partial\over \partial r }
r^2 {\partial\over \partial r }K = 0 
\eqn\laplaceex
$$
If there were no magnetic charge, then this equation would be an ordinary
Laplace equation with the usual $\delta$--function singularity
at the origin. The magnetic charges regulate the solution,
rendering it perfectly regular from the four dimensional point of view.
Physically this means there is no source at the horizon. 
Despite the curved transverse space, it 
is the flat space Laplacian that appears in \laplaceex~.
This is a remarkable property of the
metric \stringmetric~, which arises because
$$
e^{-\Phi_\perp}\sqrt{G}G^{rr}
\eqn\speccombi
$$ is equal to its flat space
value. 

\section{Classical Hair}
The classical solution is characterised by the conserved charges
at infinity. In the present context these are the $U(1)$ charges
and the ADM mass. There is a relation between these conserved charges
$$
4G_N M_{ADM} =  {\bf Q^{(1)}}+{\bf Q^{(2)}}+{\bf P^{(1)}}+{\bf P^{(2)}}
\eqn\bpssat
$$
One finds that
this saturation property is equivalent to a residual supersymmetry. 
Because it does not modify the conserved charges at infinity, hair
must therefore be consistent with residual supersymmetry.
This severely
restricts the possible form of hair for this example, 
and makes it practical
to find explicit expressions.

The following construction is inspired by the 
analogous problem of oscillations of the fundamental 
string [\callan, \waldram ]. Technically, 
the main step in 
carrying over the previous results lies in showing  
that the transverse magnetic space does
not obstruct the oscillations of the classical string.
This turns out to follow from the special property \speccombi .

First we consider adding fundamental gauge 
fields $F^{(i)}_{MN}$ that are non--trivial.
The corresponding hair will be referred to as gauge hair.  
Supersymmetry leads to the {\it ansatz}
$F^{(i)rv}\neq 0$ ~\foot{Gauge fields that are self--dual in the 
transverse space provide another possibility consistent 
with supersymmetry.  We do not consider it here. }. The equations
of motion for the gauge fields are
$$
\tilde{\nabla}_N [e^{-\Phi}F^{(i)Nv} ]= 0
\eqn\gaugehaireq
$$
Here $\tilde{\nabla}$ is the covariant derivative formed using
the generalized connection $\tilde{\Gamma}^L_{MN}=\Gamma^L_{MN}
+{1\over 2}H^L_{MN}$. In a gauge where only $A_u \neq 0$ this
reduces to 
$$
\partial_r (F^{-2}r^2 \partial_r A_u^{(i)}) = 0
\eqn\redgaugehair
$$
The calculation that shows this relies on the special property of
\speccombi . The solution is $A_u^{(I)} \propto F$ because
$F^{-1}$ is a solution to Laplace's equation in flat space.
The important qualitative point
is that the proportionality constant can be {\it any} function of $u$ .
Thus the explicit form of the gauge hair is
$$
F^{(i)}_{ru}=  {1\over {\bf Q^{(2)}}}
\partial_r F ~ q^{(i)}(u)
= {1\over (r+{\bf Q^{(2)}})^2} q^{(i)} (u)
\eqn\gaugehair
$$
The internal index $i=1,\cdots, 16$ so this hair amounts to $16$
chiral fields in $1+1$ dimensions.

The gauge hair generates energy--momentum, which must be included
as source--terms in the Einstein equations. Due to the special
form of the hair,  only one component is affected. 
This is \laplaceex~,  which is modified to read
$$
- \nabla^2 K =
{1\over (r+{\bf Q^{(2)}})^4}
\sum_{i=1}^{16} q^{(i)}(u)^2 
\eqn\uueom
$$
Here $\nabla^2$ is the Laplacian in a flat transverse space so all terms 
that depend on the magnetic charges dropped out again.
The equation is easily integrated to give
$$
K={{\bf Q^{(1)}}\over r} - {1\over 2}
{1\over r(r+{\bf Q^{(2)}})} \sum_{i=1}^{16} q^{(i)}(u)^2 
\eqn\keq
$$
The first term enters as a constant of integration that could depend on $u$.
It is determined by the condition that the charges at infinity are
unchanged. With this modification of the metric, the black hole with hair is
an exact solution. The matching condition derived in the bulk of the
paper turns out to amount to the condition that the function $K$
be regular at $r=0$. It relates the amplitudes of the
hair to the background charge. Specifically, it requires
the presence of hair.   

If there are several types of hair, it is their total that enters into
the matching condition.

In the original solution \stringmetric~ the fundamental gauge fields
$A_J^{(i)}$
were chosen to be zero, but solutions with non--zero fields are
related to the one given by T--duality.  The radial profile of
the fields obtained this way is the same as that of the hair,
so the black hole with $u$-independent `hair' 
is formally related by T--duality 
to the one without hair.
Unlike such a solution, however, the true gauge hair depends on the $u$
coordinate. The macroscopic charge is the average over the 
compactified dimension of the microscopic charge and can not 
change due to hair. 
We must therefore insist that the gauge hair has no constant mode, but all
other modes constitute acceptable hair.

We should also consider the possibility of hair that reside solely in the
metric. For hair in the longitudinal part of
the metric the most general {\it ansatz}, consistent with supersymmetry
\foot{The transverse metric might also allow non--trivial excitations,
but we will not consider that here.},
can be shown to be
$$
dS^2 = F du ( dv + K du + 2V_i dx^i ) +G_{ij}dx^i dx^j 
\eqn\hairmetric
$$
Here $K$ and $V_i$ may depend on $u$, but not on $v$. The appropriate
components of the Einstein equations are
$$
\tilde{\nabla}^i
e^{-\Phi_\perp}(\partial_i V_k -\partial_k V_i) =0
\eqn\haireom
$$
This equation allows each $V_k~;~~k=1,\cdots,8$ 
to be an arbitrary function of $u$. Following the
steps of [\waldram ], it can be shown that this solution is the
most general regular one. It does not modify any of the other
Einstein equations, so it is an exact solution. 

With a non--vanishing $V_i$ the metric is no longer asymptotically
Minkowskian. For large dimensions $V_i$ corresponds to momentum, 
while for compactified ones it corresponds to electric charge.
Since hair is not allowed to contribute to the macroscopic
quantities, the constant
mode is forbidden, but the higher modes constitute legitimate hair.
As a slightly different treatment of this hair, consider a change
to asymptotically Minkowskian coordinates 
$x^i\rightarrow x^i-\delta^{ij}V_j$.
In the new coordinate system 
the horizon is no longer at $r=0$, but has $u$--dependent origin.
We see that the hair corresponds to the breaking
of translational invariance --- it is ``Goldstone hair''.
In fact, the gauge hair can also be understood in this way, because the
fundamental gauge fields are related to the breaking
of translational invariance on the internal 16--dimensional torus. 
Taken together all the hair amounts to $24$ arbitrary chiral functions
that respect the periodicity of the $9$th coordinate. There is a
chiral bosonic string's worth of hair.

\end

\endpage


\end